\documentclass[]{article}
\usepackage{latexsym}
\usepackage{epsfig}

\begin{document}
\baselineskip=5mm
\quad

\vspace{8pt} \centerline{\Large \bf Production Ratios of Strange
Baryons from QGP with Diquarks}

\vspace{32pt} \centerline{ \large Hong Miao$^1$, Zhongbiao Ma$^1$,
Chongshou Gao$^{1,2}$}

\vspace{16pt} \centerline{ $^1$Department of Physics, Peking
University, Beijing 100871, China}

\vspace{4pt} \centerline{ $^2$Institute of Theoretical Physics,
Academia Sinica, Beijing 100080, China}

\vspace{16pt}
\begin{abstract}

Assuming that axial-vector and scalar diquarks exist in the
Quark-Gluon Plasma near the critical temperature $T_c$, baryons
can be produced through the processes of quarks and diquarks
forming $(\frac{1}{2})^+$ baryon states. Ratios of different
baryons can be estimated through this method, if such kind of QGP
with diquarks can exists.
 \vspace*{8pt}
\\ {\bf PACS number(s)\/}: 12.38.Mh 25.75.-q
\\{\bf Key words\/}: QGP, plasma, diquark, strange, baryon
\end{abstract}

\vspace{12pt}

\leftline{ \bf 1. Introduction}

\vspace{8pt}

The Quark-Gluon Plasma is generally concerned to consist of quarks
and gluons. While diquarks may exist in QGP too. They have the
color $\{3^*\}$ representation, and the flavor $\{6\}$ or
$\{3^*\}$ representations\cite{sign:sign diquark_form}. That means
an SU(3) sextet of axial-vector diquarks and an SU(3) triplet of
scalar diquarks. If axial-vector and scalar diquarks exist near
the critical temporature $T_c$ (not too high to melt
diquarks\cite{sign:sign diquark_dense_condensate}\cite{sign:sign
diquark_phase}) and approximate thermal equilibrium could form,
baryon production can be described as the process of quark and
diquark forming a $(\frac{1}{2})^+$ baryon state. Ratios of
different baryons can be estimated through this method. Since
strange baryon production is widely discussed\cite{sign:sign
Rafelski_phL}\cite{sign:sign Rafelski_Acta1}\cite{sign:sign
Rafelski_Acta2}\cite{sign:sign Rafelski_Acta3} and has a upper
limit in the hadronic gas model\cite{sign:sign Strange}, the ratio
beyond that limit can pridict that the QGP has formed or not in
the relativistic heavy ion collisions.

\vspace{12pt}

\leftline{ \bf 2. Quark-Diquark Model}

\vspace{8pt}

In the SU(6) quark-diquark model, baryon wave functions can be
written as these forms\cite{sign:sign diquark_form}\cite{sign:sign
diquark_PN}\cite{sign:sign diquark_LamdaO}\cite{sign:sign
diquark_Lamda},

\begin{eqnarray}
 \nonumber
 \mid{p}\rangle^{\uparrow}&=&\frac{1}{\sqrt{18}}[\sqrt{2}(\sqrt{2}V_{+}(uu)d^{\downarrow}-V_0(uu)d^{\uparrow})-(\sqrt{2}V_{+}(ud)u^{\downarrow}-
V_0(ud)u^{\uparrow})+3S(ud)u^{\uparrow}],\\
 \nonumber
 \mid{n}\rangle^{\uparrow}&=&\frac{1}{\sqrt{18}}[-\sqrt{2}(\sqrt{2}V_{+}(dd)u^{\downarrow}-V_0(dd)u^{\uparrow})+(\sqrt{2}V_{+}(ud)d^{\downarrow}-
V_0(ud)d^{\uparrow})+3S(ud)d^{\uparrow}],\\
 \mid{\Lambda}\rangle^{\uparrow}&=&\frac{1}{\sqrt{12}}[(\sqrt{2}V_{+}(us)d^{\downarrow}-V_0(us)d^{\uparrow})-(\sqrt{2}V_{+}(ds)u^{\downarrow}-V_0(ds)u^{\uparrow})\\
 \nonumber
 &+&S(us)d^{\uparrow}-S(ds)u^{\uparrow}+2S(ud)s^{\uparrow}],\\
 \nonumber
 \mid{\Sigma^0}\rangle^{\uparrow}
 &=&\frac{1}{\sqrt{36}}[2(\sqrt{2}V_{+}(ud)s^{\downarrow}-V_0(us)s^{\uparrow})-(\sqrt{2}V_{+}(us)d^{\downarrow}-V_0(us)d^{\uparrow})\\
 \nonumber
 &-&(\sqrt{2}V_{+}(ds)u^{\downarrow}-V_0(ds)u^{\uparrow})+3S(us)d^{\uparrow}+3S(ds)u^{\uparrow}],
\end{eqnarray}
And they can be rewritten as
\begin{eqnarray}
 \nonumber
 \mid{p}\rangle^{\uparrow}&=&\frac{1}{\sqrt{3}}[B(V_{uu},d)-\sqrt{\frac{1}{2}}B(V_{ud},u)+\sqrt{\frac{3}{2}}B(S_{ud},u)],\\
 \nonumber
 \mid{n}\rangle^{\uparrow}&=&\frac{1}{\sqrt{3}}[-B(V_{dd},u)+\sqrt{\frac{1}{2}}B(V_{ud},d)+\sqrt{\frac{3}{2}}B(S_{ud},d)],\\
 \nonumber
 \mid{\Lambda}\rangle^{\uparrow}&=&\frac{1}{\sqrt{3}}[\sqrt{\frac{3}{4}}B(V_{us},d)-\sqrt{\frac{3}{4}}B(V_{ds},u)+\sqrt{\frac{1}{4}}B(S_{us},d)-\sqrt{\frac{1}{4}}B(S_{ds},u)+B(S_{ud},s)],\\
 \nonumber
 \mid{\Sigma^0}\rangle^{\uparrow}&=&\frac{1}{\sqrt{3}}[B(V_{ud},s)-\sqrt{\frac{1}{4}}B(V_{us},d)-\sqrt{\frac{1}{4}}B(V_{ds},u)+\sqrt{\frac{3}{4}}B(S_{us},d)+\sqrt{\frac{3}{4}}B(S_{ds},u)],
\end{eqnarray}
where B represent a $(\frac{1}{2})^+$ baryon state.

So, for $p$, $n$, $\Lambda$ and $\Sigma^0$,
\begin{eqnarray}
\nonumber
 \frac{d p}{d t}&=&\frac{1}{3}\cdot[\Gamma(V_{uu},d,p)+\frac{1}{2}\Gamma(V_{ud},u,p)+\frac{3}{2}\Gamma(S_{ud},u,p)],\\
 \frac{d n}{d t}&=&\frac{1}{3}\cdot[\Gamma(V_{dd},u,n)+\frac{1}{2}\Gamma(V_{ud},d,n)+\frac{3}{2}\Gamma(S_{ud},d,n)],\\
\nonumber
 \frac{d \Lambda}{dt}&=&\frac{1}{3}\cdot[\frac{3}{4}\Gamma(V_{us},d,\Lambda^0)+\frac{3}{4}\Gamma(V_{ds},u,\Lambda^0)
 +\frac{1}{4}\Gamma(S_{us},d,\Lambda^0)+\frac{1}{4}\Gamma(S_{ds},u,\Lambda^0)+\Gamma(S_{ud},s,\Lambda^0)],\\
\nonumber
 \frac{d \Sigma^0}{dt}&=&\frac{1}{3}\cdot[\Gamma(V_{ud},s,\Sigma^0)+\frac{1}{4}\Gamma(V_{us},d,\Sigma^0)+\frac{1}{4}\Gamma(V_{ds},u,\Sigma^0)+\frac{3}{4}\Gamma(S_{us},d,\Sigma^0)+\frac{3}{4}\Gamma(S_{ds},u,\Sigma^0)],
\end{eqnarray}

Now, baryon productions can be calculated by combining different
processes of quarks and diquarks forming $(\frac{1}{2})^+$ medium
baryon states. $\Xi^0$, $\Xi^-$ and other $(\frac{1}{2})^+$
baryons can be calculated through similar methods.

\vspace{12pt}

\leftline{ \bf 3. Baryon Production with Diquarks }

\vspace{8pt}

If diquarks could exist in the Quark-Gluon Plasma near the
critical temperature, the contribution of direct 3-quark
interaction can be neglected, for the 3-body interaction has much
larger initial phase space and the interacting probability is much
smaller compared to the quark-diquark interaction. Thus, baryon
production can be described as a combination of different
processes of quarks and diquarks forming $(\frac{1}{2})^+$
baryons, as
\begin{equation}
 \frac{dB}{dt}=\sum{C_{cg}^2(D_{q_1q_2},q_3,B)\Gamma(D_{q_1q_2},q_3,B)},
\end{equation}
and $C_{cg}^2(D_{q_1q_2},q_3,B)$ is the Clebsch-Gordan coefficient
to represent the state of quark-diquark coupling shown in
equations (2), and
\begin{equation}
 \Gamma(D_{q_1q_2},q_3,B)=|v_{D_{q_1q_2}}-v_{q_3}|\rho_{D}\rho_{q_3}\sigma(D_{q_1q_2}+q_3\rightarrow{B}),
\end{equation}
where quarks and diquarks are under Fermi and Bose distributions,
\begin{equation}
 \rho_{D}=\frac{\omega_D}{\pi^2}\int_{m_D}^{\infty}\frac{p_DE_DdE_D}{e^{\frac{E_D-\mu_D}{T}}-1}
\end{equation}
\begin{equation}
 \rho_{q}=\frac{\omega_q}{\pi^2}\int_{m_q}^{\infty}\frac{p_qE_qdE_q}{e^{\frac{E_q-\mu_q}{T}}+1}
\end{equation}
\begin{eqnarray}
 \nonumber \sigma(D+q\rightarrow{B})&=&(2\pi)^4\int\frac{\omega_D\omega_q|M|^2}{4E_{D}E_q|v_{D}-v_{q}|}\\
 &&\delta^4(p_{D}+p_q-p_B)\delta(p_B^2-m_B^2)\theta(E_B)\frac{d^4p_B}{(2\pi)^3},
\end{eqnarray}

The result is\cite{sign:sign crossection}
\begin{equation}
 \frac{dB}{dt}=\sum{C_{cg}^2(D_{q_1q_2},q_3,B)\frac{3\omega_D\omega_q|M|^2}{32\pi^2}T^2F_{FB}(q_3,D_{q_1q_2},B,T)},
\end{equation}
where
\begin{equation}
 T^2F_{FB}(q,D,B,T)=\int\int\frac{dE_qdE_{D}}{(e^{\frac{E_q-\mu_q}{T}}+1)(e^{\frac{E_{D}-\mu_{D}}{T}}-1)},
\end{equation}
and the integrating ranges are
\begin{eqnarray*}
 m_q\leq{E_q}\leq{\infty}\\
 m_{D}\leq{E_{D}}\leq{\infty}
\end{eqnarray*}
and

\[E_qE_{D}\geq{\frac{1}{4m_B^2}\{4(E_q+E_{D})(m_q^2E_{D}+m_{D}^2E_q)+[m_B^2-(m_q+m_{D})^2][m_B^2-(m_q-m_{D})^2]\}}
\]

For axial-vector diquarks, one has
\begin{equation}
 L_{intV} = ig\bar{B}\gamma_{\mu}\gamma_{5}q V_{\mu},
\end{equation}

\begin{equation}
 |M|^2_V =\frac{g^2}{3}[\frac{(m_B^2-m_q^2)^2}{m_{V}^2}+m_B^2+m_q^2-2m_{V}^2+6m_Bm_q],
\end{equation}

For scalar diquarks,

\begin{equation}
 L_{intS} = ig\bar{B}q S,
\end{equation}
\begin{equation}
 |M|^2_S =
 {g^2}[(m_B+m_q)^2-m_{S}^2],
\end{equation}

Where the diquark mass is assumed as
$m_D(q_1q_2)=m_{D0}+m_{q_1}+m_{q_2}$ and  the difference of
axial-vector diquark mass and scalar diquark mass is neglected as
a simple assumption. $m_{D0}$ here is about 500-700 $MeV$. That
means the contribution of gluons to build a baryon is mainly
subjected to the diquark and the current mass $m_q$ can be used
in the calculations. $m_{D0}$ should not be smaller than the
masses of constituent quarks. Additionally, it is assumed that g
is same in these reactions.

\vspace{12pt}

\leftline{ \bf 4. Results}

\vspace{8pt}

As the quark masses have wide ranges, mean masses are used in the
calculations. (Note, the results greatly depend on the quark
masses, defferences may reach about 10 percentages if extreme
value of quark masses are used, and the lighter the quark masses
are, the higher the ratios grow). The critical temperature is set
as $T_c=170MeV$\cite{sign:sign TC_lattice1}\cite{sign:sign
TC_lattice2} for $\mu_B$ near $0MeV$. The ratios of strange
baryons to proton (or neutron) grow as the base mass of diquark
$m_{D0}$ grows, and reduce as the differences of the quark masses
of different flavors grow.  Some results are listed in table 1
below (calculation errors are less than 1\%), for $\mu_B=0MeV$,
$\mu_S=0MeV$.

\begin{center}
\begin{tabular}{|c|c|c|c|c|c|c|}
 \hline $m_{D0}$ $_{(MeV)}$         & $500$  & $550$  & $600$  & $650$  & $700$   \\
 \hline $\Lambda/p$ , $\overline{\Lambda}/\bar{p}$ $_{( \% )}$     & $49.0$ & $51.4$ & $54.4$ & $58.2$ & $63.8$  \\
 \hline $\Lambda/n$ , $\overline{\Lambda}/\bar{n}$ $_{( \% )}$     & $50.0$ & $52.6$ & $55.6$ & $59.6$ & $65.4$  \\
 \hline $\Sigma^0/p$ , $\overline{\Sigma}^0/\bar{p}$ $_{( \% )}$    & $58.1$ & $61.4$ & $66.3$ & $73.8$ & $86.1$  \\
 \hline $\Sigma^0/n$ , $\overline{\Sigma}^0/\bar{n}$ $_{( \% )}$    & $59.2$ & $62.7$ & $67.7$ & $75.6$ & $88.3$  \\
 \hline $\Lambda/(p+n)$ , $\overline{\Lambda}/(\bar{p}+\bar{n})$ $_{( \% )}$ & $24.8$ & $26.0$ & $27.5$ & $29.4$ & $32.3$  \\
 \hline
\end{tabular}
\end{center}
 \centerline{Table 1 \quad Relative yields of baryons
from QGP with diquarks at $\mu_B=0 MeV$ }

\vspace{8pt}

 When setting $\mu_B=45 MeV$ (RHIC\cite{sign:sign xunu}\cite{sign:sign
 xunuQM02}), $\mu_S=\mu_q-\mu_s$ in the plasma can be calculated by the strangeness
conservation and some results including anti-baryon/baryon ratios
are listed in table 2 below. Where $\mu_S$ is the strangeness
chemical potential, $\mu_q=\frac{1}{3}\mu_B$ is the chemical
potential of $u,d$ quark and $\mu_s$ is the chemical potential of
$s$ quark.

\begin{center}
\begin{tabular}{|c|c|c|c|c|c|c|}
\hline $m_{D0}$ $_{(MeV)}$ & $500$ & $550$ & $600$ & $650$ & $700$
\\
 \hline $\mu_S$ $_{(MeV)}$          & $20.0$ & $19.6$ & $19.2$ & $18.8$ & $18.4$  \\
 \hline $\Lambda/p$ $_{( \% )}$     & $43.6$ & $45.9$ & $48.6$ & $52.3$ & $57.6$  \\
 \hline $\Lambda/n$ $_{( \% )}$     & $44.4$ & $46.9$ & $49.8$ & $53.5$ & $59.0$  \\
 \hline $\Sigma^0/p$ $_{( \% )}$    & $51.9$ & $55.1$ & $59.7$ & $66.8$ & $78.3$  \\
 \hline $\Sigma^0/n$ $_{( \% )}$    & $52.9$ & $56.3$ & $61.0$ & $68.4$ & $80.3$  \\
 \hline $\Lambda/(p+n)$ $_{( \% )}$ & $22.0$ & $23.2$ & $24.6$ & $26.4$ & $29.1$  \\
 \hline $\overline{\Lambda}/\bar{p}$ $_{( \% )}$           & $55.1$ & $57.6$ & $60.8$ & $64.7$ & $70.7$  \\
 \hline $\overline{\Lambda}/\bar{n}$ $_{( \% )}$           & $56.2$ & $58.9$ & $62.1$ & $66.3$ & $72.6$  \\
 \hline $\overline{\Sigma}^0/\bar{p}$ $_{( \% )}$          & $64.9$ & $68.4$ & $73.5$ & $82.6$ & $94.7$  \\
 \hline $\overline{\Sigma}^0/\bar{n}$ $_{( \% )}$          & $66.2$ & $69.9$ & $75.1$ & $83.5$ & $97.1$  \\
 \hline $\overline{\Lambda}/(\bar{p}+\bar{n})$ $_{( \% )}$ & $27.8$ & $29.1$ & $30.1$ & $32.8$ & $35.8$  \\
 \hline $\bar{p}/p$ $_{( \% )}$                    & $59.5$ & $59.8$ & $60.0$ & $60.4$ & $60.7$\\
 \hline $\overline{\Lambda}/\Lambda$ $_{( \% )}$   & $75.3$ & $75.1$ & $74.9$ & $74.8$ & $74.7$\\
 \hline $\overline{\Sigma}^0/\Sigma^0$ $_{( \% )}$ & $74.4$ & $74.2$ & $73.9$ & $73.7$ & $73.4$\\
 \hline $\overline{\Xi}^0/\Xi^0$ $_{( \% )}$       & $94.6$ & $93.8$ & $93.0$ & $92.2$ & $91.5$\\
 \hline $\Xi^+/\Xi^-$ $_{( \% )}$                  & $94.6$ & $93.8$ & $93.0$ & $92.2$ & $91.5$\\
\hline
\end{tabular}
\end{center}
\centerline{Table 2 \quad Relative yields of baryons from QGP with
diquarks at $\mu_B=45 MeV$ }

\vspace{4pt}

From these calculations, one can find that the production of
$\Sigma^0$ is higher than the production of $\Lambda$. That is
because $|M|^2_V$ is larger than $|M|^2_S$ here and the $\Sigma^0$
production has the contribution of the big term
$\Gamma(V_{ud},s,\Sigma^0)$ while the $\Lambda$ production has
not.

\begin{figure}[htb]
  \centering
  \includegraphics[width=0.7\textwidth]{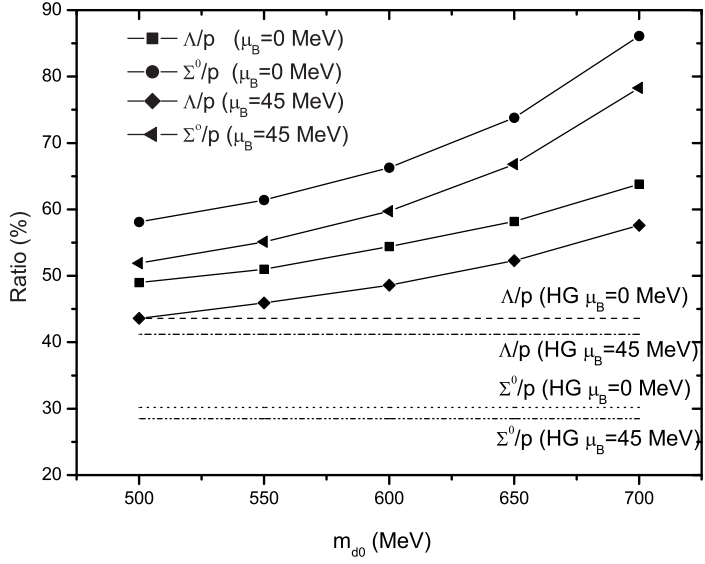}
  \caption{Ratios of $\Sigma^0$ and $\Lambda$ over proton via diquark base mass (for $mu_B=0MeV$ and $mu_B=45MeV$) and the upper limits of hadronic gas model.}
\end{figure}

\vspace{12pt}

\leftline{ \bf 5. Hadronic Gas Model}

\vspace{8pt}

For the hadronic gas of thermal and chemical equibrium, the
relative yields of hadrons can be calculated\cite{sign:sign
Strange}\cite{sign:sign crossection}\cite{sign:sign Satz}, based
on the free gas model.

The temperature of hadronic matter should be lower than the
critical temperature $T_c$ of phase transition, so the relative
yields of strange hadrons from hadronic matter of temperature
$T_c$ can give the upper limits of the relative yields of strange
hadrons\cite{sign:sign Strange}. And the production of $\Sigma^0$
is smaller than the production of $\Lambda$ due to the free gas
model.

The upper limits of some strange baryons for $T_c=170MeV$,
$\mu_B=0MeV$ are
\begin{eqnarray*}
 \frac{\Lambda}{p}<43.6\%,&&
 \frac{\Lambda}{n}<43.9\%,\\
 \frac{\Sigma^0}{p}<30.2\%,&&
 \frac{\Sigma^0}{n}<30.4\%,\\
 \frac{\Lambda}{p+n}&<&21.9\%,
\end{eqnarray*}

For $\mu_B=45 MeV$, then $\mu_S=9.8 MeV$\cite{sign:sign
Hadrongas}, the difference is small,

\begin{center}
\begin{tabular}{cccc}
 $\frac{\Lambda}{p}<41.2\%$,&
 $\frac{\Lambda}{n}<41.5\%$,&
 $\frac{\Sigma^0}{p}<28.5\%$,&
 $\frac{\Sigma^0}{n}<28.7\%$,\\
 $\frac{\overline{\Lambda}}{\bar{p}}<46.2\%$,&
 $\frac{\overline{\Lambda}}{\bar{n}}<46.5\%$,&
 $\frac{\overline{\Sigma}^0}{\bar{p}}<32.0\%$,&
 $\frac{\overline{\Sigma}^0}{\bar{n}}<32.2\%$,\\
 $\frac{\Lambda}{p+n}<20.6\%$,&
 $\frac{\overline{\Lambda}}{\bar{p}+\bar{n}}<23.2\%$,&\\
 $\frac{\overline{\Lambda}}{\Lambda}<66.1\%$\\
\end{tabular}
\end{center}

\vspace{12pt}

\leftline{ \bf 6. Discussions}

\vspace{8pt}

The ratios of strange baryons over proton or neutron produced from
QGP with diquarks are higher than the hadronic gas limits (shown
in fig.1 ). This may become a criterion to judge if QGP (with
diquarks) has been formed in the collisions and quark masses are
not too heavy. Especially, the production of $\Sigma^0$ is higher
than the production of $\Lambda$ here. Additionally,
anti-baryon/baryon ratios\cite{sign:sign xunuQM02}\cite{sign:sign
harris}\cite{sign:sign Star}\cite{sign:sign QM1} from experimental
data can be fit well by tuning $\mu_B$ and $m_{D0}$, for example,
$\mu_B=45MeV$ and $m_{D0}=650MeV$, except $\overline{\Xi}/\Xi$ is
a bit larger (exp. $0.82\pm0.08$).

Moreover, for the RHIC experiments, the baryonic potential is
quite small\cite{sign:sign xunu}\cite{sign:sign
xunuQM02}\cite{sign:sign harris}\cite{sign:sign
Star}\cite{sign:sign QM1}. The results given here are based on
this condition. But for large baryonic potentials, Bose condensate
may occurs\cite{sign:sign diquark_dense_condensate}\cite{sign:sign
diquark_phase}, and the productons of some baryons may form a peak
there, if QGP could exist at those conditions.

\vspace{12pt}

\leftline{ \bf Acknowledgement }

\vspace{8pt}

This work was supported in part by the National Natural Science
Foundation of China (90103019), and the Doctoral Programme
Foundation of Institution of Higher Education, the State Education
Commission of China (2000000147).

\end{document}